\documentclass[a4paper,12pt]{article}
\usepackage[left=3.17cm,right=3.17cm,top=2.54cm,
headheight=0.5cm,headsep=0.54cm,bottom=2.54cm,footskip=0.79cm
]{geometry}

\usepackage{amsmath,amsthm,amssymb,dsfont,cite}
\newtheorem{theorem}{Theorem}
\newtheorem{lemma}{Lemma}
\newtheorem{corollary}{Corollary}
\newtheorem{definition}{Definition}
\newtheorem{property}{Property}

\usepackage{hyperref}

\usepackage{tikz}
\usetikzlibrary{arrows}
\usetikzlibrary{decorations.markings}

\begin{document}

\title{Connes spectral distances, quantum discord and coherence of qubits}

\author{Bing-Sheng Lin$^{1, \dag}$, Zi-Hao Xu$^{1}$, Ji-Hong Wang$^{2}$, Han-Liang Chen$^{1}$\\
\small $^{1}$ School of Mathematics, South China University of Technology,
Guangzhou 510641, China\\
\small $^{2}$ Software Engineering Institute of Guangzhou, Guangzhou 510990, China\\
\small $^{\dag}$Email: sclbs@scut.edu.cn\\
}

\date{\today}

\maketitle

\begin{abstract}
We construct spectral triples of one- and two-qubit states using the Hilbert-Schmidt operatorial formulation, and study the Connes spectral distances.
We also construct the Dirac operator corresponding to the normal quantum trace distances.
Based on the Connes spectral distances, we propose some definitions of quantum discord and coherence measure of quantum states, and explicitly calculate the coherence of one-qubit states.
We also study some simple cases about two-qubit states, and the corresponding spectral distances satisfy the Pythagoras theorem.
These results are significant for studies on physical relations and geometric structures of qubits and other quantum states.
\\


\end{abstract}

\section{Introduction}
In quantum mechanics, a physical system is represented by some kind of quantum state. In order to study properties of the physical systems and also the relations between quantum states, or measure the distinguishability between the states, one can define some kinds of abstract distance measures between quantum states. For example, quantum trace distance and quantum fidelity \cite{Nielsen}.
In quantum information science, one can use quantum trace distance or quantum fidelity to quantify the differences between quantum states.

One usually consider the quantum systems with two levels (namely, qubits) in quantum information and quantum computation. To describe a quantum system with two levels, one can use the Grassmann representation of Fermi operators in fermionic phase spaces.
Since quantum phase spaces are some kinds of noncommutative spaces, one can also use mathematical tools in noncommutative geometry to study the geometric structures of quantum states in phase spaces \cite{Connes}.
In a noncommutative space, a pure state is the analog of a traditional
point in a normal commutative space, and the Connes spectral distance between
pure states corresponds to the geodesic distance between points \cite{Connes1}.
The Connes spectral distances in some kinds of noncommutative spaces have already been studied in the literature
\cite{Bimonte,Dai,Cagnache,Wallet,Martinetti1,Martinetti,DAndrea,Pythagoras,Pythagoras1,Franco,Scholtz,Chaoba,
	Revisiting,Kumar,Chakraborty,Clare,Barrett,Lin1}.
For example, Dai \emph{et. al.} have studied Connes' distance in $1D$ lattices \cite{Dai}.
Cagnache \emph{et. al.} computed Connes spectral distances between the pure states which correspond to eigenfunctions of the quantum harmonic oscillators in the Moyal plane\cite{Cagnache}.
Martinetti \emph{et. al.} obtained the spectral distance between coherent states in the so-called double Moyal plane \cite{Martinetti}.
Scholtz and his collaborators have studied the Connes spectral distances of harmonic oscillator states and also coherent states in Moyal plane and fuzzy space \cite{Scholtz,Chaoba,Revisiting}.
In the present work, we will study the Connes spectral distance between qubits which can be represented by fermionic Fock states in phase spaces.

Quantum discord and coherence are important resources in quantum information sciences \cite{Paula,Ciccarello,Baumgratz,Shao}.
Usually, one can use some distance measures, such as quantum trace distance to define the discord and coherence of quantum state $\rho$. 
Since Connes spectral distance is also a useful distance measure of quantum states, one can use the similar method to define a coherence of quantum states based on spectral distance.

This paper is organized as follows. In Sec.~\ref{sec2}, we consider the $2D$ fermionic phase space and construct a corresponding spectral triple based on the Hilbert-Schmidt operatorial formulation.
In Sec.~\ref{sec3}, we review the definition of Connes spectral distance, and derive the explicit expressions of the spectral distances between one qubits with respect to the corresponding Bloch vectors.
In Sec.~\ref{sec4}, we propose some reasonable definitions of quantum discord and coherence based on Connes spectral distance, and explicitly calculate some simple example.
In Sec.~\ref{sec5}, we construct the Dirac operators corresponding the Euclidean distances of the corresponding Bloch vectors and quantum trace distances of one-qubit states.
In Sec.~\ref{sec6}, we calculate the Connes spectral distances between some simple cases of two-qubit states.
Some discussions and conclusions are given in Sec.~\ref{sec7}.

\section{$2D$ fermionic phase space and spectral triple}\label{sec2}
Quantum bits (qubits) are some types of two-level quantum systems, which can be represented well by Grassmann algebras.
First, let us consider the simplest $2D$ fermionic phase space $(\theta_1,\theta_2)$, and the coordinate operators $\hat{\theta}_1$, $\hat{\theta}_2$ satisfy the following anticommutation relation
\begin{equation}
	\{\hat{\theta}_i,\,\hat{\theta}_j\}\equiv \hat{\theta}_i \hat{\theta}_j+\hat{\theta}_j \hat{\theta}_i=\delta_{ij}\hbar,
	\qquad i,j=1,2.
\end{equation}
The Grassmann parity of a function $f$ is denoted by
$\varepsilon(f)$, 
which is a $\mathbb{Z}_{2}$-grading ($0$ or $1$) assigned to elements in a Grassmann algebra, classifying them as either \textit{even} ($\varepsilon(f)=0$) or \textit{odd} ($\varepsilon(f)=1$). Even (odd) parity corresponds to elements that are sums of monomials with an even (odd) number of Grassmann generators $\theta_i$.
For example, $\varepsilon (\hat{\theta}_i)=1$ and
$\varepsilon (\hat{\theta}_i \hat{\theta}_j)=0$.
One can define the following annihilation and creation operators,
\begin{equation}
	\hat{f}=\frac{1}{\sqrt{2\hbar}}\left(\hat{\theta}_{1}
	+\mathrm{i}\hat{\theta}_{2}\right),\qquad
	\hat{f}^\dag=\frac{1}{\sqrt{2\hbar}}\left(\hat{\theta}_{1}
	-\mathrm{i}\hat{\theta}_{2}\right).
\end{equation}
These operators satisfy the commutation relations $\{\hat{f},\hat{f}^{\dag}\}=1$, and $\{\hat{f},\hat{f}\}=\{\hat{f}^{\dag},\hat{f}^{\dag}\}=0$.
Let $|0 \rangle$ be the vacuum
state, there are $\hat{f}|0\rangle=0$, $\hat{f}^{\dag}|0\rangle=|1\rangle$, $\hat{f}|1\rangle=|0\rangle$, $\hat{f}^{\dag}|1\rangle=0$, and $\hat{f}=|0\rangle \langle 1 |$, $\hat{f}^\dag=|1 \rangle
\langle 0 |$.
One can also use the convenient matrix representations: $|0 \rangle=(1,0)^T$,
$|1 \rangle=(0,1)^T$, where the superscript ``T'' means matrix transposition.

The fermionic phase space $(\theta_1,\theta_2)$ is some type of noncommutative space.
In general, a noncommutative space corresponds to a spectral triple $(\mathcal{A},\mathcal{H},\mathcal{D})$ \cite{Connes}, $\mathcal{A}$ is an involutive algebra acting on a Hilbert space $\mathcal{H}$ through a representation $\pi$, $\pi(a)\in L(\mathcal{H})$ with $a\in\mathcal{A}$, and the Dirac operator $\mathcal{D}$ is a self-adjoint, densely defined operator on $\mathcal{H}$ which satisfies:

1.\;$\mathcal{D}$ can be unbounded operator but $[\mathcal{D},\pi(a)]$ is bounded with $a\in\mathcal{A}$;

2.\;$\mathcal{D}$ has compact resolvent, for $\lambda\in\mathds{C}\backslash\mathds{R}$, $(\mathcal{D}-\lambda)^{-1}$  is compact when the algebra $\mathcal{A}$ is unital or $\pi(a)(\mathcal{D}-\lambda)^{-1}$ be compact if it is non-unital.

Here we will use the Hilbert-Schmidt operatorial formulation developed in Refs.~\cite{Revisiting,Formulation} to construct a spectral triple corresponding to the fermionic phase space. One can define a fermion Fock space $F=\mathrm{span}\left\{|0\rangle,|1\rangle\right\}$ and a quantum Hilbert space $Q=\mathrm{span}\left\{|i\rangle\langle j|\right\}$,
$i,j=0,1$.
In the followings, we will also denote the elements $\psi(\hat{\theta}_1,\hat{\theta}_2)$ of the quantum Hilbert space $Q$ by $|\psi)$.
A spectral triple $(\mathcal{A},\mathcal{H},\mathcal{D})$ for the $2D$ fermionic phase space $(\theta_1, \theta_2)$ can be constructed as follows,
\begin{equation}
	\mathcal{A}=Q,
	\qquad\mathcal{H}=F\otimes \mathds{C}^2,
\end{equation}
and an element $e\in \mathcal{A}$ acts on
$\Psi=\left(|\psi_1\rangle, |\psi_2\rangle\right)^T\in\mathcal{H}$
through the diagonal representation $\pi$ as
\begin{equation}
	\pi(e)\Psi
	=\left(\begin{array}{cc}
		e & 0\\
		0 & e\\
	\end{array}\right)\left(\begin{array}{c}
		|\psi_1\rangle \\
		|\psi_2\rangle \\
	\end{array}\right)
	=\left(\begin{array}{c}
		e|\psi_1\rangle \\
		e|\psi_2\rangle \\
	\end{array}\right).
\end{equation}

The geometry of the spectral triple is mainly determined by the Dirac operator. There are many different choices of the Dirac operators, and different Dirac operators correspond to different geometric structures.
In order to construct the Dirac operator for the fermionic phase space, one can consider the following extended auxiliary noncommutative space in which the coordinate operators $\hat{\Theta}_i$ and $\hat{\Lambda}_i$ satisfy the following anticommutation relations,
\begin{equation}\label{nc}
	\{\hat{\Theta}_i,\,\hat{\Theta}_j\}
	=\{\hat{\Lambda}_i,\,\hat{\Lambda}_j\}=\delta_{ij}\hbar,\qquad
	\{\hat{\Theta}_i,\,\hat{\Lambda}_j\}=\delta_{ij}\lambda,\qquad
	i,j=1,2.
\end{equation}
Here $\lambda$ is some real parameter.
It is easy to verify that, a unitary representation of the algebra (\ref{nc}) can be obtained by the following actions on the quantum Hilbert space $Q$:
\begin{equation}
	\hat{\Theta}_i|\phi)=|\hat{\theta}_i\phi),\qquad
	\hat{\Lambda}_i|\phi)
	=\frac{\lambda}{\hbar}|\hat{\theta}_i\phi)
	+(-1)^{\varepsilon(\phi)}
	\frac{\sqrt{\lambda^2-\hbar^2}}{\hbar}|\phi\,\hat{\theta}_i).
\end{equation}
One can define the following useful operators
\begin{equation}
	\hat{B}=\hat{\Lambda}_{1}+\mathrm{i}\hat{\Lambda}_{2},\qquad
	\hat{B}^\dag=\hat{\Lambda}_{1}-\mathrm{i}\hat{\Lambda}_{2}.
\end{equation}
There are
\begin{eqnarray}
	&&\hat{B}|\phi)=\lambda\sqrt{\frac{2}{\hbar}}|\hat{f}\phi)
	-{\rm i}(-1)^{\varepsilon(\phi)}
	\sqrt{\frac{2(\hbar^2-\lambda^2)}{\hbar}}|\phi\hat{f}),\nonumber\\
	&&\hat{B}^\dag|\phi)=\lambda\sqrt{\frac{2}{\hbar}}|\hat{f}^\dag\phi)
	-{\rm i}(-1)^{\varepsilon(\phi)}
	\sqrt{\frac{2(\hbar^2-\lambda^2)}{\hbar}}|\phi\hat{f}^\dag).
\end{eqnarray}
By virtue of the result in Ref.~\cite{Gayral}, one may express the Dirac operator $\mathcal{D}$ as
\begin{equation}\label{do0}
	\mathcal{D}
	=\frac{1}{\lambda}\sum_{i=1,2}\sigma_i\hat{\Lambda}_i,
\end{equation}
where $\sigma_i$'s are the Pauli matrices.
So the Dirac operator (\ref{do0}) can be written as
\begin{equation}
	\mathcal{D}
	=\frac{1}{\lambda}\left(
	\begin{array}{cc}
		0 &\hat{\Lambda}_1-\mathrm{i}\hat{\Lambda}_2 \\
		\hat{\Lambda}_1+\mathrm{i}\hat{\Lambda}_2 & 0 \\
	\end{array}\right)
	=\frac{1}{\lambda}\left(
	\begin{array}{cc}
		0 &\hat{B}^\dag \\
		\hat{B} & 0 \\
	\end{array}\right).
\end{equation}
Consider the graded commutator $[\mathcal{D},\pi(a)]_{gr}\equiv \mathcal{D}\pi(a)-(-1)^{\varepsilon(a)}\pi(a)\mathcal{D}$ with $a\in \mathcal{A}$.
After some straightforward calculations, one can obtain $[\mathcal{D},\pi(a)]_{gr}$ acting on an element $\Phi\in Q\otimes \mathds{C}^2$ as
\begin{eqnarray}
	[\mathcal{D},\pi(a)]_{gr}\Phi
	&=&[\mathcal{D},\pi(a)]_{gr}
	\left(
	\begin{array}{c}
		|\phi_1) \\
		|\phi_2) \\
	\end{array}
	\right)\nonumber\\
	&=&\frac{1}{\lambda}\left(
	\begin{array}{cc}
		0 & [\hat{B}^\dag,a]_{gr} \\{}
		[\hat{B},a]_{gr} & 0 \\
	\end{array}\right)\left(
	\begin{array}{c}
		\!|\phi_1) \\
		\!|\phi_2) \\
	\end{array}
	\!\right)\nonumber\\
	&=&\sqrt{\frac{2}{\hbar}}\left(
	\begin{array}{cc}
		0 & [\hat{f}^\dag,a]_{gr} \\{}
		[\hat{f},a]_{gr} & 0 \\
	\end{array}\right)\left(
	\begin{array}{c}
		\!|\phi_1) \\
		\!|\phi_2) \\
	\end{array}
	\!\right).
\end{eqnarray}
Therefore, one can identify the Dirac operator $\mathcal{D}$ for the $2D$ fermionic phase space $(\theta_1, \theta_2)$  as
\begin{equation}\label{do}
	\mathcal{D}=\sqrt{\frac{2}{\hbar}}\left(
	\begin{array}{cccc}
		0 & \hat{f}^\dag \\
		\hat{f} & 0 \\
	\end{array}\right).
\end{equation}
\begin{definition}
	A spectral triple $(\mathcal{A},\mathcal{H},\mathcal{D})$ for the $2D$ fermionic phase space is defined as
	\begin{equation}
		\mathcal{A}=Q,
		\qquad\mathcal{H}=F\otimes \mathds{C}^2,
		\qquad\mathcal{D}=\sqrt{\frac{2}{\hbar}}\left(
		\begin{array}{cccc}
			0 & \hat{f}^\dag \\
			\hat{f} & 0 \\
		\end{array}\right).
	\end{equation}
\end{definition}

From the above calculations, one can see that the Hilbert-Schmidt operatorial formulation is a powerful tool in studies on noncommutative geometry. One can use it to construct Dirac operators and spectral triples
through purely algebraic methods.
Here we construct a spectral triple $(\mathcal{A},\mathcal{H},\mathcal{D})$ for the one-qubit systems.
The construction (\ref{do0}) is just corresponding to the Dirac operator $D=-i\sigma^{\mu} \partial_{\mu}$ for the canonical spectral triple.
From the above procedure, one can find that the Dirac operator (\ref{do}) for the two-level systems constructed by this formulation
is somewhat unique up to some multiplicative constant. Therefore, the spectral triple constructed above is a significant model in studies on Hilbert-Schmidt operatorial formulation and noncommutative spaces.

\section{Connes spectral distances between one-qubit states}\label{sec3}
Using the Dirac operator constructed above, one can calculate the Connes spectral distances between quantum states in the fermionic phase space.
For the quantum states $\omega$ which are normal and bounded, they can be represented by density matrices $\rho$.
The action of the state $\omega$ on an element $e\in \mathcal{A}$ can be written as
\begin{equation}
	\omega(e)=\mathrm{tr}_{F}(\rho e),
\end{equation}
where $\mathrm{tr}_{F}(\cdot)$ denotes the trace over $F$.
Suppose the quantum states $\omega_1$ and $\omega_2$ correspond to the density matrices $\rho_1$ and $\rho_2$, respectively.
\begin{definition}\cite{Connes1}
	The Connes distance between the quantum states $\omega_1$ and $\omega_2$ are 
	\begin{equation}
		d(\omega_1,\omega_2)=\sup_{e\in B}|\mathrm{tr}_{F}(\rho_1 e)-\mathrm{tr}_{F}(\rho_2 e)|
		=\sup_{e\in B}|\mathrm{tr}_{F}(\Delta\rho\, e)|,
	\end{equation}
	where $\Delta\rho=\rho_1-\rho_2$.
	The set $B=\{e\in \mathcal{A}: \|[\mathcal{D},\pi(e)]\|_{op}\leqslant 1\}$,
	and $\|A\|_{op}$ is the operator norm of $A$,
	\begin{equation}
		\|A\|_{op}\equiv\sup_{\psi\in \mathcal{H},\|\psi\|=1}\|A\psi\|,\qquad
		\|A\|^2\equiv \mathrm{tr}_{F}(A^{\dag}A).
	\end{equation}
\end{definition}
The inequality $\|[\mathcal{D},\pi(e)]\|_{op}\leqslant 1$ is the so-called \emph{ball condition}.

Since Hermitian elements can give the supremum in the Connes spectral
distance functions \cite{Iochum}, one only need to consider the optimal elements $e$ being Hermitian.
Any Hermitian element $e\in \mathcal{A}$ can be expressed as the following matrix,
\begin{equation}\label{ee}
	e=\left(
	\begin{array}{cc}
		s & w^* \\
		w & t \end{array}
	\right)=\left(
	\begin{array}{cc}
		s & u-{\rm i}v \\
		u+{\rm i}v & t \end{array}
	\right),
\end{equation}
where $w=u+{\rm i}v$, and $s,t,u,v$ are real numbers.
Using the Dirac operator $\mathcal{D}$ (\ref{do}), we have
\begin{eqnarray}\label{dpe}
	[\mathcal{D},\pi(e)]&=&\sqrt{\frac{2}{\hbar}}\left(
	\begin{array}{cc}
		0 & [\hat{f}^\dag,e] \\{}
		[\hat{f},e] & 0 \\
	\end{array}\right)\nonumber\\
	&=&\sqrt{\frac{2}{\hbar}}\left(
	\begin{array}{cc}
		0 & -[\hat{f},e]^\dag \\{}
		[\hat{f},e] & 0 \\
	\end{array}\right).
\end{eqnarray}
Since $\|[\mathcal{D},\pi(e)]\|_{op}$ is just the square root of the largest eigenvalue of the matrix $[\mathcal{D},\pi(e)]^\dag[\mathcal{D},\pi(e)]$,
using the ball condition and the above matrix representations (\ref{dpe}), after some straightforward calculations, one can obtain
\begin{equation}
	2|w|^2+(s-t)^2+|s-t|\sqrt{4|w|^2+(s-t)^2}
	\leqslant \hbar.
\end{equation}
It is easy to see that, there are the following inequalities,
\begin{equation}\label{wst}
	|w|\leqslant \sqrt{\frac{\hbar}{2}},\qquad |s-t|\leqslant \sqrt{\frac{\hbar}{2}}.
\end{equation}

In general, the density matrice $\rho$ for a qubit can be expressed as \cite{Nielsen}
\begin{equation}\label{rr}
	\rho=\frac{I+\vec{r}\cdot \vec{\sigma}}{2}
	=\frac{1}{2}\left(
	\begin{array}{cc}
		1+z & x-{\rm i}y \\
		x+{\rm i}y & 1-z \end{array}
	\right),
\end{equation}
where the real vector $\vec{r}=(x, y, z)$ are the so-called Bloch vector, $|\vec{r}|\leqslant1$, and $\vec{\sigma}=(\sigma_1, \sigma_2, \sigma_3)$, $\sigma_i$ are the Pauli matrices.
Consider the states $\rho_1$ and $\rho_2$ corresponding to the Bloch vectors $\vec{r}_1=(x_1, y_1, z_1)$ and $\vec{r}_2=(x_2, y_2, z_2)$, respectively.
Using the matrix representation (\ref{ee}), one can obtain
\begin{equation}
	\mathrm{tr}_{F}(\Delta\rho\, e)=\frac{1}{2}(s-t)\Delta z+u\Delta x+v\Delta y,
\end{equation}
where
$\Delta x =x_1-x_2$, $\Delta y =y_1-y_2$, $\Delta z =z_1-z_2$.
So we have
\begin{eqnarray}
	d(\rho_1,\rho_2)&=&\sup_{e\in B}|\mathrm{tr}_{F}(\Delta\rho\, e)|
	=\sup_{e\in B}\left|\frac{1}{2}(s-t)\Delta z+u\Delta x+v\Delta y\right|\nonumber\\
	&\leqslant&\sup_{e\in B}\left(\frac{1}{2}|(s-t)\Delta z|+|u\Delta x+v\Delta y|\right)\nonumber\\
	&\leqslant&\sup_{e\in B}\left(\frac{1}{2}|(s-t)\Delta z|
	+\sqrt{u^2+v^2}\sqrt{(\Delta x)^2+(\Delta y)^2}\right)\nonumber\\
	&=&\sup_{e\in B}\left(\frac{1}{2}|(s-t)\Delta z|
	+|w|\sqrt{(\Delta x)^2+(\Delta y)^2}\right).
\end{eqnarray}
In the second inequality above, we have used the Cauchy-Schwartz inequality:
$
(a_1b_1+a_2b_2)^2\leqslant(a_1^2+a_2^2)(b_1^2+b_2^2),
$
where the equality holds if $a_1 b_2=a_2 b_1$.
Therefore, one may choose the optimal element $e$ satisfying
\begin{equation}\label{uv}
	v\Delta x=u\Delta y.
\end{equation}

For the given states $\rho_1$ and $\rho_2$, in order to attain the supremum of $|\mathrm{tr}_{F}(\Delta\rho\, e)|$, one must choose the Hermitian element $e$ to make $|s-t|$ and $|w|$ as large as possible. So there should be
\begin{equation}
	2|w|^2+(s-t)^2+|s-t|\sqrt{4|w|^2+(s-t)^2}=\hbar,
\end{equation}
and
\begin{equation}
	|s-t|=\frac{1}{\sqrt{2\hbar}}(\hbar-2|w|^2).
\end{equation}
Therefore, we have
\begin{eqnarray}
	|\mathrm{tr}_{F}(\Delta\rho\, e)|
	&\leqslant&\frac{1}{2}|(s-t)\Delta z|
	+|w|\sqrt{(\Delta x)^2+(\Delta y)^2}\nonumber\\
	&=&\frac{1}{2}\sqrt{\frac{\hbar}{2}}|\Delta z|-\frac{|\Delta z|}{\sqrt{2\hbar}}|w|^2
	+|w|\sqrt{(\Delta x)^2+(\Delta y)^2}\nonumber\\
	&=&\frac{1}{2}\sqrt{\frac{\hbar}{2}}\frac{|\Delta \vec{r}|^2}{|\Delta z|}
	-\frac{|\Delta z|}{\sqrt{2\hbar}}\left(|w|
	-\sqrt{\frac{\hbar}{2}}\frac{\sqrt{(\Delta x)^2+(\Delta y)^2}}{|\Delta z|}\right)^2\nonumber\\
	&\leqslant&\frac{1}{2}\sqrt{\frac{\hbar}{2}}\frac{|\Delta \vec{r}|^2}{|\Delta z|},
\end{eqnarray}
where $\Delta \vec{r}=\vec{r}_1-\vec{r}_2=(\Delta x, \Delta y, \Delta z)$, $|\Delta \vec{r}|=\sqrt{(\Delta x)^2+(\Delta y)^2+(\Delta z)^2}\leqslant 2$.

When
$(\Delta x)^2+(\Delta y)^2\leqslant (\Delta z)^2$, from Eq.~(\ref{wst}), 
one can choose the optimal element $e$ satisfying the relation (\ref{uv}) and
\begin{equation}\label{ww}
	|w|
	=\frac{\sqrt{(\Delta x)^2+(\Delta y)^2}}{|\Delta z|}\sqrt{\frac{\hbar}{2}}.
\end{equation}
Therefore, we have
\begin{equation}
	u=\sqrt{\frac{\hbar}{2}}\frac{\Delta x}{\Delta z},\quad v=\sqrt{\frac{\hbar}{2}}\frac{\Delta y}{\Delta z},\quad
	|s-t|=\sqrt{\frac{\hbar}{2}}\frac{(\Delta z)^2-(\Delta x)^2-(\Delta y)^2}{(\Delta z)^2}.
\end{equation}
For simplicity, one can choose $s>0$ and $t=0$, then we have
\begin{equation}
	s=\sqrt{\frac{\hbar}{2}}\frac{(\Delta z)^2-(\Delta x)^2-(\Delta y)^2}{(\Delta z)^2},
\end{equation}
and the optimal element $e$ can be chosen as
\begin{equation}
	e=\sqrt{\frac{\hbar}{2}}\left(
	\begin{array}{cc}
		\frac{(\Delta z)^2-(\Delta x)^2-(\Delta y)^2}{(\Delta z)^2} & \frac{\Delta x-{\rm i}\Delta y}{\Delta z} \\[0.5em]
		\frac{\Delta x+{\rm i}\Delta y}{\Delta z} & 0 \end{array}
	\right).
\end{equation}
In this case, there is
\begin{equation}\label{sd1}
	d(\rho_1,\rho_2)
	=\frac{1}{2}\sqrt{\frac{\hbar}{2}}\frac{(\Delta x)^2+(\Delta y)^2+(\Delta z)^2}{|\Delta z|}
	=\frac{1}{2}\sqrt{\frac{\hbar}{2}}\frac{|\Delta \vec{r}|^2}{|\Delta z|}.
\end{equation}

When $(\Delta x)^2+(\Delta y)^2> (\Delta z)^2$,
one can only choose the optimal element $e$ satisfying the relation (\ref{uv}) and
\begin{equation}
	|w|
	=\sqrt{\frac{\hbar}{2}}<\frac{\sqrt{(\Delta x)^2+(\Delta y)^2}}{|\Delta z|}\sqrt{\frac{\hbar}{2}},\qquad
	|s-t|=\frac{1}{\sqrt{2\hbar}}(\hbar-2|w|^2)=0.
\end{equation}
For simplicity, one can choose $s=t=0$.
From the relation (\ref{uv}), we have
\begin{equation}
	u=\sqrt{\frac{\hbar}{2}}\frac{\Delta x}{\sqrt{(\Delta x)^2+(\Delta y)^2}},\quad v=\sqrt{\frac{\hbar}{2}}\frac{\Delta y}{\sqrt{(\Delta x)^2+(\Delta y)^2}}.
\end{equation}
The optimal element $e$ can be chosen as
\begin{equation}
	e=\sqrt{\frac{\hbar}{2}}\left(
	\begin{array}{cc}
		0 & \frac{\Delta x-{\rm i}\Delta y}{\sqrt{(\Delta x)^2+(\Delta y)^2}} \\[0.5em]
		\frac{\Delta x+{\rm i}\Delta y}{\sqrt{(\Delta x)^2+(\Delta y)^2}} & 0 \end{array}
	\right).
\end{equation}
In this case, there is
\begin{equation}\label{sd2}
	d(\rho_1,\rho_2)
	=\frac{1}{2}|(s-t)\Delta z|
	+|w|\sqrt{(\Delta x)^2+(\Delta y)^2}
	=\sqrt{\frac{\hbar}{2}}\sqrt{(\Delta x)^2+(\Delta y)^2}.
\end{equation}

It is convenient to use the spherical coordinates in the Bloch sphere, one can also denote $\Delta \vec{r}=\vec{r}_1-\vec{r}_2=r(\sin\theta\cos\phi, \sin\theta\sin\phi, \cos\theta)$, where $\theta$ is the polar angle from the positive $z$-axis with
$0\leqslant\theta\leqslant \pi$, $\phi$ is the azimuthal angle in the $xy$-plane from the $x$-axis with $0\leqslant\phi< 2\pi$, and $r$ is the distance from the end point of $\vec{r}_1$ to the end point of $\vec{r}_2$ with $0\leqslant r\leqslant 2$.

\begin{theorem}
	In the spectral triple $(\mathcal{A},\mathcal{H},\mathcal{D})$, the Connes spectral distance between one-qubit states $\rho_1$ and $\rho_2$ is
	\begin{equation}\label{drr}
		d(\rho_1, \rho_2)=
		\left\{
		\begin{array}{ll}
			\sqrt{\dfrac{\hbar}{2}}\sqrt{(\Delta x)^2+(\Delta y)^2}
			=r\sin\theta\sqrt{\dfrac{\hbar}{2}},
			& \qquad\hbox{$\dfrac{\pi}{4}\leqslant \theta \leqslant \dfrac{3\pi}{4}$;} \\[1em]
			\dfrac{1}{2}\sqrt{\dfrac{\hbar}{2}}\dfrac{|\Delta \vec{r}|^2}{|\Delta z|}
			=\dfrac{r}{2|\cos\theta|}\sqrt{\dfrac{\hbar}{2}},
			& \qquad\hbox{others.}
		\end{array}
		\right.
	\end{equation}
\end{theorem}

\begin{proof}
It is easy to see that, the vector $\Delta \vec{r}$ is in the ball with $0\leqslant |\Delta \vec{r}|= r \leqslant 2$, and
\begin{equation}
	\sin \theta=\frac{\sqrt{(\Delta x)^2+(\Delta y)^2}}{\sqrt{(\Delta x)^2+(\Delta y)^2+(\Delta z)^2}}
	=\frac{\sqrt{(\Delta x)^2+(\Delta y)^2}}{r},\qquad
	\cos \theta=\frac{\Delta z}{r}.
\end{equation}

When
$(\Delta x)^2+(\Delta y)^2\leqslant (\Delta z)^2$, there is
$0\leqslant \sin\theta \leqslant\dfrac{\sqrt{2}}{2}$,
and $0\leqslant \theta \leqslant \dfrac{\pi}{4}$ or $\dfrac{3\pi}{4}\leqslant \theta \leqslant \pi$. From Eq.~(\ref{sd1}), we have
\begin{equation}
	d(\rho_1,\rho_2)
	=\frac{1}{2}\sqrt{\frac{\hbar}{2}}\frac{|\Delta \vec{r}|^2}{|\Delta z|}
	=\dfrac{r}{2|\cos\theta|}\sqrt{\dfrac{\hbar}{2}}.
\end{equation}

When $(\Delta x)^2+(\Delta y)^2\geqslant (\Delta z)^2$,
there is
$\dfrac{\sqrt{2}}{2}\leqslant \sin\theta \leqslant 1$,
and $\dfrac{\pi}{4}\leqslant \theta \leqslant\dfrac{3\pi}{4}$. From Eq.~(\ref{sd2}), we have
\begin{equation}
	d(\rho_1,\rho_2)
	=\sqrt{\frac{\hbar}{2}}\sqrt{(\Delta x)^2+(\Delta y)^2}
	=r\sin\theta\sqrt{\dfrac{\hbar}{2}}.
\end{equation}
\end{proof}

These results are similar to those in Refs.~\cite{Cagnache,Pythagoras1}, but the spectral triple constructed in the present work is different from those considered in Refs.~\cite{Cagnache,Pythagoras1}. The method used in the present work is also different from those used in the literature.

From the result (\ref{drr}), one can find that these spectral distances are additive when the corresponding points of the states in the Bloch sphere are collinear,
\begin{equation}
	d(\rho_1, \rho_3)=d(\rho_1, \rho_2)+d(\rho_2, \rho_3).
\end{equation}
Furthermore, when the corresponding points of the states in the Bloch sphere are
on the same horizontal plane, namely $\Delta z=0$, the Connes spectral distances between the states are proportional to the Euclidean distances in the Bloch representation with a factor $\sqrt{\hbar/2}$.
It is easy to see that, for the diagonal states, namely $x_i=y_i=0$ in (\ref{rr}), the optimal elements $e$ for the Connes spectral distances can also be diagonal, this is similar to the result in Ref.~\cite{Lin2}.

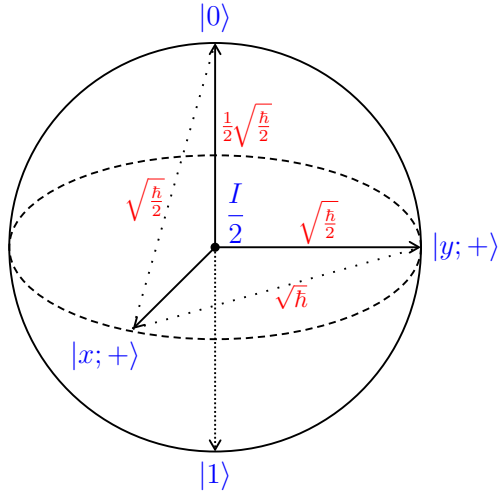
\begin{figure}
	\centering
	\begin{tikzpicture}[>=angle 60,thick,baseline=0pt]
		\coordinate(o)at(0,0);
		\draw(o)circle(3cm);
		\draw[fill](o)circle(1.5pt) node[blue] at (0.3,0.5) {$\dfrac{I}{2}$};
		\draw[->](o)--(-1.2,-1.2) node[blue] at (-1.6,-1.6) {$|x;+\rangle$};
		\draw[->](o)--(3,0)node[blue,right]{$|y;+\rangle$} node[red] at (1.5,0.37) {\footnotesize $\sqrt{\frac{\hbar}{2}}$};
		\draw[->](o)--(0,3)node[blue,above]{$|0\rangle$} node[red] at (0.45,1.8) {\footnotesize $\frac{1}{2}\!\sqrt{\frac{\hbar}{2}}$};
		\draw[rotate around={0.:(0.,0.)},densely dashed](0,0)ellipse(3cm and 1.35cm);
		\draw[densely dotted,->](o)--(0,-3)node[blue,below]{$|1\rangle$};
		\draw[loosely dotted](-1.2,-1.2)--(0,3) node[red] at (-1.03,0.7) {\footnotesize $\sqrt{\frac{\hbar}{2}}$};
		\draw[loosely dotted](-1.2,-1.2)--(3,0) node[red] at (1.1,-0.73) {\footnotesize $\sqrt{\hbar}$};
	\end{tikzpicture}
	\caption{\label{fig1}Connes spectral distances between one-qubit states in the Bloch sphere.}
\end{figure}

Using the formulas (\ref{drr}), one can obtain the Connes spectral distances between any one-qubit states. Some results are depicted in Fig.~\ref{fig1}.

\section{Connes spectral distance measure of quantum discord and quantum coherence}\label{sec4}

Quantum trace distance is an important mathematical tool in quantum information sciences \cite{Nielsen}.
One can use it to analyse the relations of quantum states.
The quantum trace distances between qubits are equal to half the corresponding Euclidean distances in the Bloch representations. In general, the results of Connes spectral distances are quite different from those about the quantum trace distances of one-qubit states.
Similar to the trace distance, one can use the Connes spectral distance to analyse the physical properties and relations of quantum states, such as quantum discord and coherence.

Quantum discord is an important information-theoretic measure
of nonclassical correlations \cite{Paula,Ciccarello}.
Consider a composite quantum system $\mathcal{H}_{AB}$, which consists of two subsystems $A$ and $B$. 
The state $\rho_{CQ}\in\mathcal{H}_{AB}$ is a classical-quantum state if and only if it can be written as the following form \cite{Paula}
\begin{equation}
	\rho_{CQ} = \sum_{k} p_k \Pi_k^A \otimes \rho_k^B,
\end{equation}
with $0 \leq p_k \leq 1$ ($\sum_k p_k = 1$), $\{\Pi_k^A\}$ denoting a set of orthogonal projectors for subsystem $A$, and $\rho_k^B$ being a general reduced density operator for subsystem $B$.
The quantum discord reaches zero for and only for the classical-quantum
state. So we can look at the quantum discord as some type of `distance' between the state $\rho$ and the set of classical-quantum states.
For example, one can define the trace distance discord of the state $\rho$ as follow \cite{Ciccarello}
\begin{equation}
	D_{tr}(\rho)=\min_{\chi\in S_{CQ}}d_{tr}(\rho,\chi),
\end{equation}
where $S_{CQ}$ is the set of classical-quantum states, and $d_{tr}(\rho,\chi)$ is the quantum trace distance between the quantum states $\rho$ and $\chi$. 

Since Connes spectral distance is also a distance measure of quantum states, it is reasonable to define a spectral distance discord of a state $\rho$ based on Connes spectral distance as follow.
\begin{definition}
	A spectral distance discord of a quantum state $\rho$ is defined as
	\begin{equation}
		D_{SD}(\rho)=\min_{\chi\in S_{CQ}}d(\rho,\chi),
	\end{equation}
	where $S_{CQ}$ is the set of classical-quantum states, and $d(\rho,\chi)$ is the Connes spectral distance between the quantum states $\rho$ and $\chi$. 
\end{definition}

Quantum coherence is also an important resource in quantum information sciences \cite{Baumgratz}.
In the formalism presented
in Ref.~\cite{Baumgratz}, a non-negative convex
function $C$ defined on the  space of states $\rho$ is called a coherence measure if it
satisfies the following two conditions:
\begin{enumerate}
	\item Monotonicity under incoherent channel $\Lambda^I$: \[C(\Lambda^I[\rho])\leq C(\rho),\]
	\item Strong monotonicity under incoherent channel $\Lambda^I$:
	\begin{equation}
		\sum_np_nC(\rho_n)\leq C(\rho),
	\end{equation}
\end{enumerate}
where $\rho_n:=(K_n\rho K_n^\dagger)/p_n$, $p_n:=\mathrm{tr}(K_n\rho K_n^\dagger)$, $K_n$'s are incoherent  Kraus operators satisfying $\sum_nK_n^\dagger K_n=\mathbb{I}$.

Usually, one can use some distance measures, such as quantum trace distance\cite{Baumgratz,Shao} to define the coherence of quantum state $\rho$. For example,
a coherence of a quantum state $\rho$ can be defined as
\begin{equation}
	C_{tr}(\rho)=\min_{\phi\in\mathcal{I}}d_{tr}(\rho,\phi),
\end{equation}
where $\mathcal{I}$ is the set of incoherent states, and $d_{tr}(\rho,\phi)$ is the quantum trace distance between the quantum states $\rho$ and $\phi$. 
The elements in $\mathcal{I}$ are just the diagonal states in a fixed basis.

Therefore, it is also reasonable to define a coherence of a state $\rho$ based on Connes spectral distance as follow.
\begin{definition}
	A coherence of a quantum state $\rho$ is defined as
	\begin{equation}\label{sdc}
		C_{SD}(\rho)=\min_{\phi\in\mathcal{I}}d(\rho,\phi),
	\end{equation}
	where $\mathcal{I}$ is the set of incoherent states, and $d(\rho,\phi)$ is the Connes spectral distance between the quantum states $\rho$ and $\phi$. 
\end{definition}

As an example, let us explicitly calculate the coherence of arbitrary one-qubit states using the above definition (\ref{sdc}).
The given state $\rho$ and any incoherent state $\phi$ can be express as
\begin{equation}
	\rho=\frac{1}{2}\left(
	\begin{array}{cc}
		1+z & x-{\rm i}y \\
		x+{\rm i}y & 1-z \end{array}
	\right),
\qquad
	\phi
	=\frac{1}{2}\left(
	\begin{array}{cc}
		1+z' & 0 \\
		0 & 1-z' \end{array}
	\right),
\end{equation}
and
\begin{equation}
	\Delta\rho=\rho-\phi=\frac{1}{2}\left(
	\begin{array}{cc}
		\Delta z & x-{\rm i}y \\
		x+{\rm i}y & -\Delta z \end{array}
	\right),
\end{equation}
where $\Delta z=z-z'$.
The Bloch vectors of $\rho$ and $\phi$ are $\vec{r}=(x, y, z)$ and $\vec{r}\,'=(0, 0, z')$, respectively, and $\Delta\vec{r}=(x, y, \Delta z)$.

For the given state $\rho$ and $(x, y, z)$, if
$x^2+y^2\leqslant (\Delta z)^2$, there is
\begin{equation}
	d(\rho,\phi)
	=\frac{1}{2}\sqrt{\frac{\hbar}{2}}\frac{|\Delta \vec{r}|^2}{|\Delta z|}
	=\frac{1}{2}\sqrt{\frac{\hbar}{2}}\frac{x^2+y^2+(\Delta z)^2}{|\Delta z|}
	\geqslant\sqrt{\frac{\hbar}{2}}\sqrt{x^2+y^2}.
\end{equation}
If $x^2+y^2\geqslant (\Delta z)^2$,
there is
\begin{equation}
	d(\rho,\phi)
	=\sqrt{\frac{\hbar}{2}}\sqrt{x^2+y^2}.
\end{equation}
Therefore, for the one-qubit state $\rho$, the coherence is
\begin{equation}
	C_{SD}(\rho)=\min_{\phi\in\mathcal{I}}d(\rho,\phi)=\sqrt{\frac{\hbar}{2}}\sqrt{x^2+y^2},
\end{equation}
and the nearest incoherent state can be chosen as $\rho_{\textrm{diag}}$ with the Bloch vector $\vec{r}\,'=(0, 0, z)$.
This is similar to the results of the $l_1$ norm of coherence \cite{Baumgratz} and also the trace norm of coherence defined in Ref.~\cite{Shao}.

\section{The Dirac operators for Euclidean distances and quantum trace distances}\label{sec5}
The quantum trace distance is equal to half the Euclidean distance between the corresponding Bloch vectors of quantum states.
Since the geometric structures of the spectral triples are mainly determined by the Dirac operators, it is interesting to consider the Dirac operators corresponding to the spectral distance $d(\rho_1, \rho_2)$ which equals the normal Euclidean distance between the corresponding Bloch vectors of quantum states, namely, $d(\rho_1, \rho_2) =|\vec{r}_1-\vec{r}_2|=|\Delta\vec{r}|= \sqrt{(\Delta x)^2 + (\Delta y)^2 + (\Delta z)^2}$.

Regarding that the Dirac operator (\ref{do}) has the following matrix representation,
\begin{equation}
	\mathcal{D}=\sqrt{\frac{2}{\hbar}}\left(
	\begin{array}{cccc}
		0 & \hat{f}^\dag \\
		\hat{f} & 0 \\
	\end{array}\right)=\frac{1}{\sqrt{2\hbar}}( \sigma_1\otimes \sigma _{1}  + \sigma_2\otimes \sigma _{2} ),
\end{equation}
one can consider the following deformed Dirac operator
\begin{equation}
	\mathcal{D}_{E} = c ( \sigma_1\otimes \sigma _{1}  + \sigma_2\otimes \sigma _{2})  + \delta\sigma_3\otimes \sigma _{3}, 
\end{equation}
where $c$ and $\delta$ are non-zero real numbers.

Using the expression (\ref{ee}), after some straightforward calculations, one can obtain the eigenvalues of $[ \mathcal{D}_{E}, \pi(e) ]^{\dagger} [ \mathcal{D}_{E}, \pi(e) ]$ as follows,
\begin{equation}
	\left( \sqrt{c^2 (s - t)^2 + 4 \delta^2 |w|^2} \pm c \sqrt{(s - t)^2 + 4
		|w|^2}  \right)^2.
\end{equation}
So we have
\begin{equation}
	\| [ \mathcal{D}_{E}, \pi(e) ] \|_{op} 
	= \sqrt{c^2 (s - t)^2 + 4 \delta^2 |w|^2} + |c| \sqrt{(s - t)^2 + 4|w|^2}.
\end{equation}
Using the ball condition $\| [ \mathcal{D}_{E}, \pi(e) ] \|_{op} \leqslant 1$, one can obtain the following constraint relation
\begin{equation}\label{wst1}
	\sqrt{c^2 (s - t)^2 + 4 \delta^2 |w|^2} + |c| \sqrt{(s - t)^2 + 4|w|^2}\leqslant 1.
\end{equation}
Obviously, there are
\begin{equation}\label{stw}
	|s - t|\leqslant \frac{1}{2|c|}, \qquad |w|\leqslant\frac{1}{2(|c|+|\delta|)}.
\end{equation}
The spectral distance between quantum states $\rho_1$ and $\rho_2$ is
\begin{eqnarray}
	d_E(\rho_1, \rho_2)&=&\sup_{e\in B}|\mathrm{tr}_{F}(\Delta\rho\, e)|
	=\sup_{e\in B}\left|\frac{1}{2}(s-t)\Delta z+u\Delta x+v\Delta y\right|\nonumber\\
	&\leqslant& \sup_{e \in B}\left(\sqrt{\frac{1}{4}(s-t)^{2} +u^{2}+v^{2}   } \sqrt{(\Delta x)^2 + (\Delta y)^2+(\Delta z)^2}\right)\nonumber\\
	&=&\sup_{e \in B}\left(\sqrt{\frac{1}{4}(s-t)^{2} +\left | w  \right |^{2}     }\  |\Delta\vec{r}|\right).
\end{eqnarray}
In the inequality above, we have used the Cauchy-Schwartz inequality,
and the equality holds if
\begin{equation}\label{uvst}
	\frac{\Delta x}{u}=\frac{\Delta y}{v}=\frac{\Delta z}{\frac{1}{2}(s - t)}.
\end{equation}

For the given states $\rho_1$ and $\rho_2$, in order to attain the supremum of $|\mathrm{tr}_{F}(\Delta\rho\, e)|$, one must choose the optimal elements $e$ to make $|s-t|$ and $|w|$ as large as possible. From the condition (\ref{wst1}), there should be
\begin{equation}
	\sqrt{c^2 (s - t)^2 + 4 \delta^2 |w|^2} + |c| \sqrt{(s - t)^2 + 4|w|^2}= 1.
\end{equation}
Therefore, we have
\begin{equation}
	|s-t|=\frac{1}{2|c|}\sqrt{16 (c^2 - \delta^2)^2 |w|^4 - 8 (c^2 + \delta^2) |w|^2 + 1}.
\end{equation}
Note that, under the condition (\ref{stw}), there is always
\begin{equation}
	16 (c^2 - \delta^2)^2 |w|^4 - 8 (c^2 + \delta^2) |w|^2 + 1\geqslant 0	.
\end{equation}
So we have
\begin{eqnarray}\label{dd}
	d_E(\rho_1,\rho_2)&\leqslant&\sup_{e\in B}\left(\sqrt{\frac{1}{16 c^2}\big(16 (c^2 - \delta^2)^2 |w|^4 - 8 (c^2 + \delta^2) |w|^2 + 1\big)+|w|^2}\ 
	|\Delta\vec{r}|\right)\nonumber\\
	&=&\sup_{e\in B}\left(\frac{1}{4 |c|} \big[1+ 4 (c^2 - \delta^2) |w|^2\big] 
	|\Delta\vec{r}|\right).
\end{eqnarray}

From the relation (\ref{uvst}), for different states $\rho_1$, $\rho_2$, namely, different $\Delta x$, $\Delta y$, $\Delta z$, one should choose different optimal element $e$ with different $w$.
So from (\ref{dd}), it is easy to see that, for any states $\rho_1$, $\rho_2$, 
in order to obtain $d_E(\rho_1,\rho_2)=|\Delta\vec{r}|$,
there should be $c^2- \delta^2=0$, or $|c|=|\delta|$, and
\begin{equation}
	\frac{1}{4 |c|} \big[1+4 (c^2 - \delta^2) |w|^2\big]=\frac{1}{4 |c|}=1.
\end{equation}
For simplicity, one may choose $c,\delta>0$, and there are
\begin{equation}
	c=\delta=\frac{1}{4}.
\end{equation}
\begin{lemma}
	The Dirac operator for the Euclidean distance is
	\begin{equation}
		\mathcal{D}_{E} = \frac{1}{4} \sum_{i=1}^{3} \sigma_i\otimes \sigma _{i}  = \frac{1}{4} \left ( \sigma_1\otimes \sigma _{1}  + \sigma_2\otimes \sigma _{2}  + \sigma_3\otimes \sigma _{3} \right ), 
	\end{equation}
	where $\sigma_i$ are the Pauli matrices.
\end{lemma}

To construct the corresponding optimal elements $e$, we now have the following conditions,
\begin{equation}
	|w|=\sqrt{u^{2}+v^2}\leqslant 1, \qquad |s-t|=2\sqrt{1- |w|^2}\leqslant 2.
\end{equation}
For simplicity, one may choose the real number $t=-s$, and there are
\begin{equation}
	\left\{
	\begin{array}{ll}
		\dfrac{\Delta x}{u}=\dfrac{\Delta y}{v}=\dfrac{\Delta z}{s},\\[0.8em]
		u^{2}+v^2 +s^{2}=1,
	\end{array}\right.
\end{equation}
and the optimal element is
\begin{equation}
	e=\left(
	\begin{array}{cc}
		s & u-{\rm i}v \\
		u+{\rm i}v & -s \end{array}
	\right)=\frac{1}{|\Delta\vec{r}|}\left(
	\begin{array}{cc}
		\Delta z & \Delta x-{\rm i}\Delta y \\
		\Delta x+{\rm i}\Delta y & -\Delta z \end{array}
	\right).
\end{equation}
From the expression (\ref{rr}), we also have
\begin{equation}
	e=2\frac{\rho_1-\rho_2}{|\vec{r}_1-\vec{r}_2|}=\frac{\Delta\vec{r}}{|\Delta\vec{r}|}\cdot \vec{\sigma}.
\end{equation}
With the optimal element above, one can obtain the main results.
\begin{theorem}
	In the spectral triple $(\mathcal{A},\mathcal{H},\mathcal{D}_E)$, the Connes spectral distance between one-qubit states $\rho_1$ and $\rho_2$ is equal to the Euclidean distance between the Bloch vectors, 
	\begin{equation}
		d_E(\rho_1, \rho_2) =|\vec{r}_1-\vec{r}_2|,
	\end{equation}
	where $\vec{r}_1$, $\vec{r}_2$ are the corresponding Bloch vectors of $\rho_1$, $\rho_2$, respectively.
\end{theorem}

Since for qubit states, the quantum trace distance is equal to half the Euclidean distance between the corresponding vectors in the Bloch sphere, one can also construct the Dirac operator corresponding to the normal quantum trace distance as follow.
\begin{corollary}
	The Dirac operator for the quantum trace distance is
	\begin{equation}
		\mathcal{D}_{T} = \frac{1}{2} \sum_{i=1}^{3} \sigma_i\otimes \sigma _{i}  = \frac{1}{2} \left ( \sigma_1\otimes \sigma _{1}  + \sigma_2\otimes \sigma _{2}  + \sigma_3\otimes \sigma _{3} \right ). 
	\end{equation}
\end{corollary}

\section{Connes spectral distances between two-qubit states}\label{sec6}
Next, let us consider the Connes spectral distances between two-qubit states.
Similarly, one can construct the following Fock space
\begin{equation}
	\mathcal{F}=\mathrm{span}\left\{|m,n\rangle\equiv|m\rangle_{\!1}\otimes|n\rangle_{\!2},
	\quad m,n=0,1\right\},
\end{equation}
where $\{|0 \rangle_{\!i}, |1 \rangle_{\!i}\}$, $i=1,2$ are just the bases of $2D$ fermion Fock space in the previous section.
One may define the operators $\hat{f}_1	=|0\rangle \langle 1 |\otimes I_2$, $\hat{f}_2=I_1\otimes|0\rangle \langle 1 |$. The creation and annihilation operators $\hat{f}_i^\dag, \hat{f}_i$ satisfy the commutation relations $\{\hat{f}_i,\hat{f}_i^{\dag}\}=1$, $\{\hat{f}_i,\hat{f}_i\}=\{\hat{f}_i^{\dag},\hat{f}_i^{\dag}\}=0$, and $[\hat{f}_1,\hat{f}_2^{\dag}]=[\hat{f}_1^{\dag},\hat{f}_2]=[\hat{f}_1,\hat{f}_2]=[\hat{f}_1^{\dag},\hat{f}_2^{\dag}]=0$.
The corresponding quantum Hilbert space is constructed as follow,
\begin{equation}
	\mathcal{Q}=\mathrm{span}\left\{|m_1,n_1\rangle\langle m_2,n_2|,
	\quad m_1,n_1,m_2,n_2=0,1\right\}.
\end{equation}

Using the methods similar to those in the $2D$ case in the previous section, one can construct a spectral triple $(\mathcal{A}',\mathcal{H}',\mathcal{D}')$ as follows.
\begin{definition}
	A spectral triple $(\mathcal{A}',\mathcal{H}',\mathcal{D}')$ for the $4D$ fermionic phase space is defined as
	\begin{equation}\label{do2}
		\mathcal{A}'=\mathcal{Q},
		\qquad\mathcal{H}'=\mathcal{F}\otimes \mathds{C}^4,
		\qquad
		\mathcal{D}'=\mathrm{i}\sqrt{\frac{2}{\hbar}}\left(
		\begin{array}{cccc}
			0 & 0 & \hat{f}_2^\dag & \hat{f}_1^\dag \\
			0 & 0 &  \hat{f}_1 & -\hat{f}_2 \\
			-\hat{f}_2 & -\hat{f}_1^\dag & 0 & 0 \\
			-\hat{f}_1 & \hat{f}_2^\dag & 0 & 0 \\
		\end{array}\right).
	\end{equation}
\end{definition}

Now let us study the Connes spectral distances between two-qubit states in this noncommutative space.
Usually, the calculations in the $4D$ fermionic phase space are much more cumbersome and complicated than those in the $2D$ case.
For simplicity, here we only study the spectral distances between the states $|ij\rangle$, $i,j=0,1$.

First, let us consider the spectral distance between $|00\rangle$ and $|10\rangle$,
\begin{eqnarray}\label{d0010}
	d(|00\rangle, |10\rangle)&=&\sup_{e\in B}|\mathrm{tr}_\mathcal{F}(\rho_{00} e)-\mathrm{tr}_\mathcal{F}(\rho_{10} e)|\nonumber\\
	&=&\sup_{e\in B}|\langle 00|e|00\rangle-\langle 10|e|10\rangle|\nonumber\\
	&=&\sup_{e\in B}|\langle 00|e\hat{f}_1|10\rangle-\langle 00|\hat{f}_1 e|10\rangle|\nonumber\\
	&=&\sup_{e\in B}|\langle 00|[\hat{f}_1,e]|10\rangle|\nonumber\\
	&\leqslant&\big\|[\hat{f}_1,e]\big\|_{op}.
\end{eqnarray}
Here we have used the Bessel's inequality:
\begin{theorem}\cite{Revisiting}
	For any operator $A$ with the matrix elements $A_{ij}$ in some orthonormal bases, there is
	\begin{equation}\label{bi}
		|A_{ij}|^2\leqslant\sum_i|A_{ij}|^2\leqslant\|A\|_{op}^2.
	\end{equation}
\end{theorem}
For a Hermitian element $e\in \mathcal{A}'$,
using the above Dirac operator $\mathcal{D}'$, one can calculate the commutator $[\mathcal{D}',\pi(e)]$,
\begin{equation}
	[\mathcal{D}',\pi(e)]
	=\mathrm{i}\sqrt{\frac{2}{\hbar}}\left(
	\begin{array}{cc}
		0 & M \\
		M^\dag & 0 \\
	\end{array}\right),
\end{equation}
where
\begin{equation}\label{d1d}
	M=\left(
	\begin{array}{cc}
		-[\hat{f}_2,e]^\dag & -[\hat{f}_1,e]^\dag \\[0.2em]
		[\hat{f}_1,e] & -[\hat{f}_2,e] \\
	\end{array}\right).
\end{equation}
Since $\|M^\dag M\|_{op}=\|M M^\dag\|_{op}$, for any Hermitian element $e$, using the ball condition, one can obtain the following inequality,
\begin{equation}
	\big\|[\mathcal{D}',\pi(e)]\big\|_{op}^2
	=\left\|[\mathcal{D}',\pi(e)]^\dag[\mathcal{D}',\pi(e)]\right\|_{op}
	=\frac{2}{\hbar}\|M M^\dag\|_{op}\leqslant 1.
\end{equation}
From the above expression (\ref{d1d}), we have
\begin{equation}\label{dd2}
	M M^\dag
	=\left(
	\begin{array}{cc}
		[\hat{f}_1,e]^\dag[\hat{f}_1,e]+[\hat{f}_2,e]^\dag[\hat{f}_2,e] & ~~[\hat{f}_1,e]^\dag[\hat{f}_2,e]^\dag-[\hat{f}_2,e]^\dag[\hat{f}_1,e]^\dag \\[0.3em]
		-[\hat{f}_1,e][\hat{f}_2,e]+[\hat{f}_2,e][\hat{f}_1,e] & [\hat{f}_1,e][\hat{f}_1,e]^\dag+[\hat{f}_2,e][\hat{f}_2,e]^\dag \\
	\end{array}\right).
\end{equation}
By virtue of the Bessel's inequality (\ref{bi}), one can obtain
\begin{equation}
	\sup_{\phi\in\mathcal{F},\langle\phi|\phi\rangle=1}
	\!\!\!\!\!\!\langle\phi|[\hat{f}_1,e]^\dag[\hat{f}_1,e]+[\hat{f}_2,e]^\dag[\hat{f}_2,e]|\phi\rangle
	\leqslant\|M M^\dag\|_{op} \leqslant\frac{\hbar}{2}.
\end{equation}
Since $\langle\phi|[\hat{f}_1,e]^\dag[\hat{f}_1,e]|\phi\rangle \geqslant 0$, $\langle\phi|[\hat{f}_2,e]^\dag[\hat{f}_2,e]|\phi\rangle \geqslant 0$, we also have
\begin{equation}
	\sup_{\phi\in\mathcal{F},\langle\phi|\phi\rangle=1}
	\!\!\!\!\!\!\langle\phi|[\hat{f}_1,e]^\dag[\hat{f}_1,e]|\phi\rangle
	\leqslant \frac{\hbar}{2},\qquad
	\sup_{\phi\in\mathcal{F},\langle\phi|\phi\rangle=1}
	\!\!\!\!\!\!\langle\phi|[\hat{f}_2,e]^\dag[\hat{f}_2,e]|\phi\rangle
	\leqslant \frac{\hbar}{2},
\end{equation}
or
\begin{equation}
	\big\|[\hat{f}_1,e]\big\|_{op}^2\leqslant \frac{\hbar}{2},\qquad
	\big\|[\hat{f}_2,e]\big\|_{op}^2\leqslant \frac{\hbar}{2}.
\end{equation}
Similarly, we also have the following other inequalities.
\begin{property}
	For any Hermitian element $e$ satisfying the ball condition, there are the following inequality,
	\begin{eqnarray}\label{bc2}
		&&\sup_{\phi\in\mathcal{F},\langle\phi|\phi\rangle=1}
		\!\!\!\!\!\!\langle\phi|[\hat{f}_1,e]^\dag[\hat{f}_1,e]+[\hat{f}_2,e]^\dag[\hat{f}_2,e]|\phi\rangle
		\leqslant\frac{\hbar}{2},
		\nonumber\\
		&&\sup_{\phi\in\mathcal{F},\langle\phi|\phi\rangle=1}
		\!\!\!\!\!\!\langle\phi|[\hat{f}_1,e][\hat{f}_1,e]^\dag+[\hat{f}_2,e][\hat{f}_2,e]^\dag|\phi\rangle
		\leqslant\frac{\hbar}{2},
		\nonumber\\
		&&\sup_{\phi\in\mathcal{F},\langle\phi|\phi\rangle=1}
		\!\!\!\!\!\!\langle\phi|[\hat{f}_1,e][\hat{f}_1,e]^\dag+[\hat{f}_2,e]^\dag[\hat{f}_2,e]|\phi\rangle
		\leqslant\frac{\hbar}{2},\nonumber\\
		&&\sup_{\phi\in\mathcal{F},\langle\phi|\phi\rangle=1}
		\!\!\!\!\!\!\langle\phi|[\hat{f}_1,e]^\dag[\hat{f}_1,e]+[\hat{f}_2,e][\hat{f}_2,e]^\dag|\phi\rangle
		\leqslant\frac{\hbar}{2},
	\end{eqnarray}
	and
	\begin{equation}\label{f1e}
		\big\|[\hat{f}_1,e]\big\|_{op}^2=\big\|[\hat{f}_1^\dag,e]\big\|_{op}^2\leqslant \frac{\hbar}{2},\qquad
		\big\|[\hat{f}_2,e]\big\|_{op}^2=\big\|[\hat{f}_2^\dag,e]\big\|_{op}^2\leqslant \frac{\hbar}{2}.
	\end{equation}
\end{property}

Combine the results (\ref{d0010}) and (\ref{f1e}), there must be
\begin{equation}
	d(|00\rangle, |10\rangle)=\sup_{e\in B}|\mathrm{tr}_\mathcal{F}(\rho_{00} e)-\mathrm{tr}_\mathcal{F}(\rho_{10} e)|\leqslant \sqrt{\frac{\hbar}{2}}.
\end{equation}

Now, we only need to find some optimal elements $e$ which can saturate the above inequality.
The existence of such optimal elements $e$ means that the spectral distance $d(|00\rangle, |10\rangle)$ should be equal to $\sqrt{\hbar/2}$.

Similar to the result in Ref.~\cite{Lin2}, one can firstly consider the elements $e_o$ being diagonal,
$e_o=diag(e_1, e_2, e_3, e_4)$, where $e_i$ are real numbers.
Now the Connes spectral distance between the states $|00 \rangle$ and $|10 \rangle$ can be expressed as
\begin{eqnarray}
	d(|00\rangle, |10\rangle)&=&
	\sup_{e\in B}|\langle 00|e|00\rangle-\langle 10|e|10\rangle|\nonumber\\
	&=&\sup_{e\in B}|e_{1}-e_3|.
\end{eqnarray}

From the relations (\ref{bc2}), using the matrix expression (\ref{d1d}), after some straightforward calculations, one can obtain the following relations,
\begin{eqnarray}\label{ee12}
	&&(e_1-e_2)^2\leqslant \frac{\hbar}{2},\quad
	(e_1-e_3)^2\leqslant \frac{\hbar}{2},\quad
	(e_2-e_4)^2\leqslant \frac{\hbar}{2},\quad
	(e_3-e_4)^2\leqslant \frac{\hbar}{2},
	\nonumber\\
	&&\Big(|e_1-e_2-e_3+e_4|+\sqrt{(e_1-e_4)^2+(e_2-e_3)^2}\Big)^2\leqslant \hbar.
\end{eqnarray}
Using the above relations (\ref{ee12}), one can choose $e_1=e_2=\sqrt{\frac{\hbar}{2}}$, $e_3=e_4=0$, and obtain the following optimal element
\begin{equation}
	e_o^{(1)}
	=\left( \begin{array}{cc}
		\sqrt{\frac{\hbar}{2}} & 0 \\ 0 & 0 \end{array} \right)\otimes \mathds{I}_2,
\end{equation}
where $\mathds{I}_2$ is the $2\times 2$ identity matrix.
Finally, we obtain
\begin{equation}
	d(|00 \rangle,|10 \rangle)=|\langle 00|e_o^{(1)}|00\rangle-\langle 10|e_o^{(1)}|10\rangle|=\sqrt{\frac{\hbar}{2}}.
\end{equation}
Similarly, there is
\begin{equation}
	d(|01 \rangle,|11 \rangle)=\sqrt{\frac{\hbar}{2}},
\end{equation}
and $e_o^{(1)}$ can still be the corresponding optimal element.

Using the same method, one can obtain
\begin{equation}
	d(|00\rangle, |01\rangle)=
	d(|10\rangle, |11\rangle)=\sqrt{\frac{\hbar}{2}},
\end{equation}
and the corresponding optimal element can be chosen as
\begin{equation}
	e_o^{(2)}
	=\mathds{I}_2\otimes\left( \begin{array}{cc}
		\sqrt{\frac{\hbar}{2}} & 0 \\ 0 & 0 \end{array} \right).
\end{equation}
It is easy to verify that, there are
\begin{equation}
	[\hat{f}_1,e_o^{(2)}]=0,\qquad [\hat{f}_2,e_o^{(1)}]=0.
\end{equation}

Next, let us consider the spectral distance between $|00\rangle$ and $|11\rangle$,
\begin{eqnarray}\label{d0011}
	d(|00\rangle, |11\rangle)&=&\sup_{e\in B}|\mathrm{tr}_\mathcal{F}(\rho_{00} e)-\mathrm{tr}_\mathcal{F}(\rho_{11} e)|\nonumber\\
	&=&\sup_{e\in B}|\langle 00|e|00\rangle-\langle 11|e|11\rangle|\nonumber\\
	&=&\sup_{e\in B}|\langle 00|e|00\rangle-\langle 10|e|10\rangle+\langle 10|e|10\rangle-\langle 11|e|11\rangle|\nonumber\\
	&=&\sup_{e\in B}|\langle 00|e\hat{f}_1|10\rangle-\langle 00|\hat{f}_1 e|10\rangle+\langle 10|e\hat{f}_2|11\rangle-\langle 10|\hat{f}_2 e|11\rangle|\nonumber\\
	&=&\sup_{e\in B}|\langle 00|[\hat{f}_1,e]|10\rangle+\langle 10|[\hat{f}_2,e]|11\rangle|\nonumber\\
	&\leqslant&\sup_{e\in B}\sqrt{2}\sqrt{|\langle 00|[\hat{f}_1,e]|10\rangle|^2+|\langle 10|[\hat{f}_2,e]|11\rangle|^2}.
\end{eqnarray}
In the inequality above, we have also used the Cauchy-Schwartz inequality.

Since for any states $|\phi\rangle,|\varphi\rangle$, there is $|\langle \phi|[\hat{f}_i,e]|\varphi\rangle|^2\geqslant 0$. It is easy to see that,
\begin{eqnarray}
	&&|\langle 00|[\hat{f}_1,e]|10\rangle|^2
	\nonumber\\
	&\leqslant&|\langle 00|[\hat{f}_1,e]|10\rangle|^2+|\langle 01|[\hat{f}_1,e]|10\rangle|^2+|\langle 10|[\hat{f}_1,e]|10\rangle|^2+|\langle 11|[\hat{f}_1,e]|10\rangle|^2\nonumber\\
	&=&\langle 10|[\hat{f}_1,e]^\dag[\hat{f}_1,e]|10\rangle.
\end{eqnarray}
Here we have used the resolution of the identity:
$\sum_{i,j=0,1}|ij\rangle \langle ij|=I$.
Similarly, there is
\begin{equation}
	|\langle 10|[\hat{f}_2,e]|11\rangle|^2=|\langle 11|[\hat{f}_2,e]^\dag|10\rangle|^2
	\leqslant\langle 10|[\hat{f}_2,e][\hat{f}_2,e]^\dag|10\rangle.
\end{equation}
By virtue of the inequalities (\ref{bc2}), one can obtain
\begin{eqnarray}\label{f1010}
	&&|\langle 00|[\hat{f}_1,e]|10\rangle|^2+|\langle 10|[\hat{f}_2,e]|11\rangle|^2
	\nonumber\\
	&\leqslant&\langle 10|[\hat{f}_1,e]^\dag[\hat{f}_1,e]|10\rangle+\langle 10|[\hat{f}_2,e][\hat{f}_2,e]^\dag|10\rangle\nonumber\\
	&=&\langle 10|\big([\hat{f}_1,e]^\dag[\hat{f}_1,e]+[\hat{f}_2,e][\hat{f}_2,e]^\dag\big)|10\rangle\nonumber\\
	&\leqslant&\!\!\!\!\sup_{\phi\in\mathcal{F},\langle\phi|\phi\rangle=1}
	\!\!\!\!\!\!\langle\phi|[\hat{f}_1,e]^\dag[\hat{f}_1,e]+[\hat{f}_2,e][\hat{f}_2,e]^\dag|\phi\rangle
	\leqslant\frac{\hbar}{2}.
\end{eqnarray}
Combine the results (\ref{d0011}) and (\ref{f1010}), we have
\begin{equation}
	d(|00\rangle, |11\rangle)
	=\sup_{e\in B}|\mathrm{tr}_\mathcal{F}(\rho_{00} e)-\mathrm{tr}_\mathcal{F}(\rho_{11} e)|
	\leqslant \sqrt{\hbar}.
\end{equation}
It is easy to check that, one can choose the following optimal element
\begin{equation}
	e_o
	=\frac{1}{\sqrt{2}}\big(e_o^{(1)}+e_o^{(2)}\big),
\end{equation}
and then obtain
\begin{equation}
	d(|00 \rangle,|11 \rangle)=|\langle 00|e_o|00\rangle-\langle 11|e_o|11\rangle|=\sqrt{\hbar}.
\end{equation}
Similarly, there is
\begin{equation}
	d(|01 \rangle,|10 \rangle)=\sqrt{\hbar},
\end{equation}
and the corresponding optimal element can be chosen as
\begin{equation}
	e'_o
	=\frac{1}{\sqrt{2}}\big(e_o^{(1)}-e_o^{(2)}\big).
\end{equation}

\begin{figure}
	\centering
	\begin{tikzpicture}[thick,baseline=0pt]
		\node[blue] (a) at (0,0) {$|00 \rangle$};
		\node[blue] (b) at (0,5) {$|01 \rangle$};
		\node[blue] (c) at (5,0) {$|10 \rangle$};
		\node[blue] (d) at (5,5) {$|11 \rangle$};
		\draw[dashed] (a) to node[red,left] {$\sqrt{\frac{\hbar}{2}}$} (b);
		\draw[dashed] (a) to node[red,below] {$\sqrt{\frac{\hbar}{2}}$} (c);
		\draw[dashed] (d) to node[red,above] {$\sqrt{\frac{\hbar}{2}}$} (b);
		\draw[dashed] (d) to node[red,right] {$\sqrt{\frac{\hbar}{2}}$} (c);
		\draw[dashed] (a) to node[red,near end,above left] {$\sqrt{\hbar}$} (d);
		\draw[dashed] (b) to node[red,near end,below left] {$\sqrt{\hbar}$} (c);
	\end{tikzpicture}
	\caption{\label{fig2}Connes spectral distances between two-qubit states $|ij \rangle$.}
\end{figure}
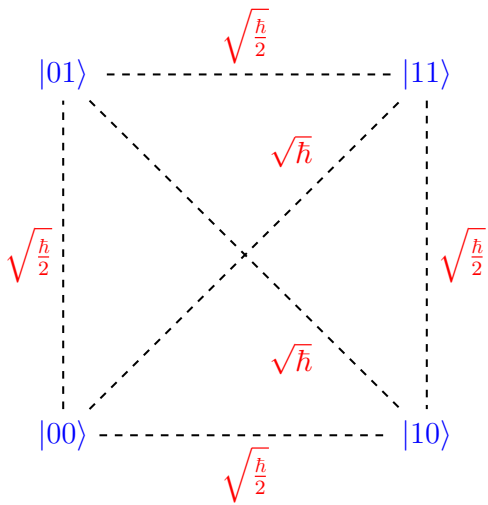

These distances are depicted in Fig.~\ref{fig2}.
From Fig.~\ref{fig2}, one can see that these spectral distances satisfy the Pythagoras theorem.
This is similar to the result in Refs.~\cite{Pythagoras,Pythagoras1}.

As a comparison, let us calculate the quantum trace distances $d_T$ between these states,
\begin{equation}
	d_T(\rho_1,\rho_2)=\frac{1}{2}\mathrm{tr}|\rho_1-\rho_2|,
\end{equation}
where $|A|:=\sqrt{A^\dag A}$.
After some straightforward calculations, one can obtain
\begin{equation}
	d_T(|ij \rangle,|kl \rangle)=1,\qquad i\neq k ~~\mathrm{and/or}~~ j\neq l.
\end{equation}

We find that the Connes spectral distances between these two-qubit states are quite different from their trace distances. One can also use these Connes spectral distances to measure the relations between qubits. In some special cases, it should give some new different results than the quantum trace distances.
For example, from the perspective of normal quantum trace distance, the relationships between the above states $|ij \rangle$ can be regarded as the same. But we can further distinguish these states to some extent using Connes spectral distances.
In this sense, the Connes spectral distance can be consider as a useful supplement to the quantum trace distances in quantum information sciences.

Similarly, one can using the above methods to study Connes spectral distances between $n$-qubit states in higher-dimensional noncommutative spaces.

\section{Discussions and conclusions}\label{sec7}
In this paper, we study the Connes spectral distances between one- and two-qubit states which can be represented by some fermionic Fock states.
By virtue of the Hilbert-Schmidt operatorial formulation, we construct a spectral triple corresponding to the $2D$ fermionic phase spaces, and calculate the Connes spectral distance between fermionic Fock states.

We also construct the Dirac operators and spectral triples in which the Connes spectral distances equal the Euclidean distances of the corresponding Bloch vectors or quantum trace distances of one-qubit states.
Furthermore, we study some simple cases about two-qubit states. In these simple cases, the spectral distances satisfy the Pythagoras theorem. These results are significant for the study of physical relations and geometric structures of qubits and other quantum states.

From the above results of the Connes spectral distances between qubits, one can find that the Connes spectral distances are usually quite different from quantum trace distances. One can use Connes spectral distances to measure the relations between quantum states. For example, one can use these different distance measures to study the discord and coherence of qubits. In some special cases, it should give some new different results.

So we believe that the Connes spectral distances can be used as a significant supplement to the quantum trace distances in quantum information sciences.
As an application, 
we have used the Connes spectral distance to define quantum discord and coherence of qubit states, and explicitly calculate the coherence of one-qubit states.
In some simple cases, it can obtain similar results as the $l_1$ norm of coherence and also the trace norm of coherence defined in the literature.

Furthermore, one can also study the spectral distances between other kinds of pure states and mixed states in higher-dimensional noncommutative spaces.
But the calculations will be much more cumbersome and complicated.
Our method used in the present work is different from those used in the literature.
We hope that our methods and results can help the studies of mathematical structures of noncommutative spaces and also physical properties of quantum systems.

\section*{Acknowledgements}
This article is dedicated to Professor Ke Wu in Capital Normal University in celebration of his 80th birthday.
The authors would like to thank the anonymous referees for careful reading and valuable comments. This work is partly supported by the Guangdong Basic and Applied Basic Research Foundation (Grant No. 2024A1515010380), the 2024 Guangdong Province Education Science Planning Project (Higher Education Special) (No. 2024GXJK455), the 2024 Guangdong Higher Education Teaching Reform Project.

\end{document}